\documentclass[%
 prd, twocolumn, 
 amsmath,amssymb,
 aps, physrev,
]{revtex4-2}

\usepackage{graphicx}
\usepackage{dcolumn}
\usepackage{bm}

\usepackage{mathtools}
\usepackage{amsmath}
\usepackage{amssymb}
\usepackage{braket}
\usepackage{color}
\usepackage{xcolor}
\usepackage{hyperref}
\usepackage{slashed}
\usepackage{graphicx}
\usepackage{mathrsfs}
\usepackage{tikz}
\usepackage{tikz-cd}
\usepackage{enumitem}
\usepackage{eso-pic}
\usepackage{subfigure}

\usepackage{mathtools} 
\usepackage{extarrows} 
\usepackage{multirow}

\newcommand{\newreptheorem}[2]{%
\newenvironment{rep#1}[1]{
 \def\rep@title{#2 \ref{##1}}%
 \begin{rep@theorem}}%
 {\end{rep@theorem}}}
\makeatother

\newreptheorem{lemma}{Lemma}

\newreptheorem{conj}{Conjecture}

\newcommand{\be}{\begin{equation}}
\newcommand{\ee}{\end{equation}}
\newcommand{\ba}{\begin{aligned}}
\newcommand{\ea}{\end{aligned}}
\newcommand{\bea}{\begin{eqnarray}}
\newcommand{\eea}{\end{eqnarray}}

\def\tr{\mathop{\mathrm{tr}}\nolimits}

\newcommand{\reportnum}[2]{
  \AddToShipoutPictureBG*{%
    \AtPageUpperLeft{%
      \hspace{0.75\paperwidth}%
      \raisebox{#1\baselineskip}{%
        \makebox[0pt][l]{\textnormal{#2}}
  }}}%
}

\def\Ehalf{E_{7+1/2}}

\def\mb{\mathbb}
\def\mbf{\mathbf}
\def\mc{\mathcal}

\def\bp{\begin{pmatrix}}
\def\ep{\end{pmatrix}}
\def\ptl{\partial}

\def\t{\widetilde}

\def\gYM{g_{\mathrm{YM}}}

\begin{document}

\reportnum{-3}{USTC-ICTS/PCFT-26-08}

\title{\textbf{On $E_{7+1/2}$ gauge theory} 
}%

\author{Xin Wang$^{1,2}$}
\author{Yi-Nan Wang$^{3,4,2}$}
\affiliation{$^1$Interdisciplinary Center for Theoretical Study,
University of Science and Technology of China}
\affiliation{$^2$Peng Huanwu Center for Fundamental Theory, Hefei, Anhui 230026, China}

\affiliation{%
 $^3$School of Physics, Peking University 
}%

\affiliation{$^4$Center for High Energy Physics, Peking University}


\date{\today}

\begin{abstract}
We study gauge theory based on the intermediate Lie algebra $E_{7+1/2}$, interpolating between $E_7$ and $E_8$. We propose a concrete UV completion via a 6d SCFT whose tensor branch description contains a pure $E_{7+1/2}$ gauge sector. The proposal is tested by 6d anomaly cancellation and by the 5d $\mathcal N=1$ Coulomb branch prepotential from the associated M-theory geometry. As a nonperturbative check, we determine the elliptic genus of the single-string worldsheet CFT using modular bootstrap. The result matches the vacuum character of the corresponding VOA for $E_{7+1/2}$, completing the elliptic genus/VOA correspondence along the Deligne--Cvitanovi\'{c} series.

\end{abstract}

\maketitle



\textit{Introduction.} 
\textit{Gauge theory} provides the fundamental theoretical framework for the description of elementary particles. Its central physical object is the gauge field, which takes value in the Lie algebra and transforms under the gauge group. In the existing literature about gauge theory, the gauge algebra is typically taken as a direct sum of a semisimple Lie algebra with copies of $U(1)$, i.e. a \textit{reductive Lie algebra}. Nonetheless, a particular class of non-reductive Lie algebras, named \textit{intermediate Lie algebra}, appears in the recent study of 2d conformal field theories~\cite{Lee:2023owa,Lee:2024fxa,Sun:2024mfz,Lee:2026ngm} and the vertex operator algebra of 4d and 3d $\mc{N}=2$ superconformal field theories~\cite{Kim:2024dxu,Deb:2025ypl,Deb:2025ddc}. One particular important example is the 190 dimensional $\Ehalf$ Lie algebra, which lies in between the exceptional Lie algebras $E_7$ and $E_8$ in the Deligne–Cvitanovi\'{c} series ($E_7\subset\Ehalf\subset E_8$)~\cite{Cvitanovic:2008zz,deligne1996serie}.

In this letter, we propose that $\Ehalf$ can be realized as an effective gauge symmetry in six-dimensional superconformal field theories (SCFTs) engineered in F-theory \cite{Morrison:1996na,Morrison:1996pp,Morrison:2012np,Morrison:2012js}. The theory admits a tensor-branch description that contains a \emph{pure} $\Ehalf$ gauge sector~\footnote{This possibility was conjectured in \cite{Lee:2023owa}.}. We test this proposal in three complementary ways: (i) we verify its consistency with six-dimensional anomaly cancellation; (ii) we match the five-dimensional low-energy effective theory obtained from the dual M-theory geometry; and (iii) we compute a protected BPS-string observable—the one-string elliptic genus—whose result agrees with the predicted vacuum character of a
vertex operator algebra (VOA) and the associated generalized Schur index in lower spacetime dimensions \cite{Beem:2013sza,DelZotto:2016pvm,Beem:2017ooy,Deb:2025ypl,Lee:2023owa}. Together, these checks provide evidence for the existence of an intermediate gauge algebra within the 6d SCFT landscape and open a new avenue for exploring a largely missing corner of 6d and lower-dimensional SCFTs via dimensional reductions.

\textit{$\Ehalf$ Lie algebra.} We first present the definition of $\Ehalf$ Lie algebra~\cite{Sextonions}. Recall that the adjoint representation $\mbf{248}$ of $E_8$ has the branching rule under the subalgebra $E_7\times U(1)\subset E_8$: 
\be
\label{E7U1-branching}
\mbf{248}\rightarrow\mbf{133}_0+\mbf{56}_1+\mbf{56}_{-1}+\mbf{1}_2+\mbf{1}_0+\mbf{1}_{-2}\,.
\ee
The generators $\mbf{133}_0+\mbf{56}_1+\mbf{1}_2$ form a closed Lie algebra, which is the 190-dimensional $\Ehalf$. In terms of the branching rule of $E_7\subset \Ehalf$, the adjoint representation $\mbf{190}$ of $\Ehalf$ is thus decomposed into $\mbf{190}\rightarrow\mbf{133}+\mbf{56}+\mbf{1}$. The Lie bracket of these generators preserve the $U(1)$ grading in (\ref{E7U1-branching}). The Lie bracket between two elements of the $\mathbf{133}$ part is given by the standard $E_7$ Lie bracket. The Lie bracket between one element $X\in \mathbf{133}$ and one element $h\in\mathbf{56}$ is given by the standard Lie algebra action of $E_7$ on the representation $\mbf{56}$:
\be
[X,h]:=X\cdot h\,.
\ee
The Lie bracket between two elements $g,h\in \mathbf{56}$ is given by the symplectic form $\omega(g,h)\in\mb{C}$:
\be
\label{omega}
[g,h]=\omega(g,h)t\,,
\ee
where $t$ is the generator for the $\mathbf{1}$ part of $\Ehalf$. $\omega$ preserves the $E_7$ action:
\be
\label{E7-omega}
\omega(u,X\cdot v)=-\omega(X\cdot u,v)\,
\ee
for $u,v\in\mbf{56}$. Finally, $t$ commutes with all the generators of $\Ehalf$. The Jacobi identity of the Lie brackets straightforwardly follows from the embedding of $\Ehalf\subset E_8$.

$\Ehalf$ is not a semisimple Lie algebra, and we cannot define its root and weight systems. When we mention the root and weights of $\Ehalf$, they always mean the corresponding ones under the subalgebra $E_7\subset\Ehalf$. For example, for the Coulomb branch of $\Ehalf$ all physical fields are charged under the Cartan subalgebra $U(1)^7\subset E_7\subset\Ehalf$.

For matter fields charged under $\Ehalf$, the irreducible representations of $\Ehalf$ was discussed in e.g. \cite{Lee:2023owa}. For the simplest non-trivial representation $\mbf{57}$, in terms of the branching rule (\ref{E7U1-branching}) it is decomposed into $\mbf{56}_{-1}+\mbf{1}_0$. We decompose the 57-dimensional vector space of $\mbf{57}$ into $(v,z)$, where $v$ is a 56-dimensional vector that is the fundamental representation of $E_7$, and $z\in\mb{C}$ is a complex number. The action of $\mbf{190}=\mbf{133}_0+\mbf{56}_1+\mbf{1}_2$ generators on $\mbf{57}=\mbf{56}_{-1}+\mbf{1}_0$ should respect the $U(1)$ grading. The action of the Lie algebra elements $X\in \mbf{133}$, $u\in \mbf{56}$, $t\in\mbf{1}$ of $\Ehalf$ takes the form of
\be
\ba
\label{57-action}
X\cdot (v,z)&=(X\cdot v,0)\cr
u\cdot (v,z)&=(0,\omega(u,v))\cr
t\cdot (v,z)&=(0,0)\,.
\ea
\ee
One can check the consistency condition of Lie algebra action
\be
[g,h]\cdot (v,z)=g\cdot (h\cdot (v,z))-h\cdot (g\cdot (v,z))
\ee
for $g,h\in\Ehalf$, given the identity (\ref{E7-omega}). The $\mbf{57}$ representation is not faithful, as the center element $t$ acts trivially on it.

A subtlety arises in the bilinear symmetric form $g_{ab}=\tr(t_a t_b)$ for the $\Ehalf$ generators $t_a$, $t_b$. From (\ref{57-action}) and the Lie bracket, it can be checked that $g_{ab}$ is only non-vanishing if $t_a,t_b\in\mbf{133}$. The degeneration of $g_{ab}$ has an important consequence in the interpretation of $\Ehalf$ gauge theory, which will be explained later.

\textit{6d pure $\Ehalf$ gauge theory.} Having presented the definition of $\Ehalf$ Lie algebra, we now propose that the $\Ehalf$ gauge theory frequently appears in the landscape of 6d (1,0) supergravity~\cite{Kumar:2009ac,Kumar:2010ru,Morrison:2012np,Morrison:2012js} and SCFTs~\cite{HeckmanMorrisonVafa,Heckman:2015bfa,Heckman:2018jxk}.

\begin{figure}
\subfigure[]{%
    \includegraphics[width=0.3\linewidth]{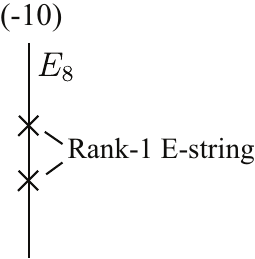}
    \label{fig:E-string}
  }\hfill
\subfigure[]{%
    \includegraphics[width=0.19\linewidth]{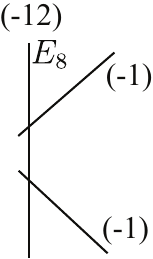}
    \label{fig:12-E8}
  }\hfill
  \subfigure[]{%
    \includegraphics[width=0.12\linewidth]{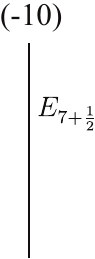}
    \label{fig:E7half}
  }\hfill

\caption{Three descriptions of a $(-10)$-curve in 6d F-theory with generic elliptic fibration over it: (a) $E_8$ gauge theory coupled to two copies of rank-1 E-string on a $(-10)$-curve; (b) Blow up $(-10)$-curve twice, to get $E_8$ gauge theory on a $(-12)$-curve; (c) $\Ehalf$ gauge theory on a $(-10)$-curve.}\label{fig:10}

\end{figure}

We consider F-theory on a singular elliptic threefold $X_3$, which is a generic fibration of a complex surface $B_2$. The local complex coordinates near a single $(-10)$-curve \footnote{The $(-10)$-curve is schematically a two-sphere $S^2$ with self-intersection number $-10$. This singular configuration was conjectured and briefly commented on in \cite{Lee:2023owa}.} $\Sigma:u=0$ are denoted by $(u,v)$. The Weierstrass model of the elliptic CY3 $X_3$ is given by~\cite{Morrison:2012np}
\be
\label{10-Weierstrass}
y^2=x^3+f_4(v) u^4 x z^4+(1+av+bv^2)u^5 z^6+\dots\,,
\ee
where higher order terms in $u$ are omitted. The order of vanishing of $(f,g)$ on $u=0$ equals $(4,5)$, with two codimension-two (4,6) loci at $u=1+av+bv^2=0$, where $a,b$ are two complex parameters. In F-theory literature, the gauge theory on $\Sigma$ was conventionally interpreted as an $E_8$ gauge theory coupled to two copies of rank-1 E-string~\cite{Morrison:1996pp,Morrison:2012np}, which are strongly coupled 6d (1,0) SCFTs, see figure~\ref{fig:E-string}.

To analyze the 6d gravitational anomaly, we embed the local geometry (\ref{10-Weierstrass}) in a compact elliptic Calabi-Yau threefold, that is the generic fibration over the Hirzebruch surface $\mb{F}_{10}$. In this above interpretation we can check the 6d gravitational anomaly cancellation~\cite{Randjbar-Daemi:1985tdc,Sagnotti:1992qw}
\be
\label{grav-anomaly}
273-29T=H_{\rm neutral}+H_{\rm charged}-V\,.
\ee
As the numbers of tensor multiplets, vector multiplets, neutral hypermultiplets (under the Cartan subalgebra) are $T=1$, $V=248$, $H_{\rm neutral}=h^{2,1}(\widetilde{X}_3)+1=434$~\cite{Morrison:2012js}, where $\widetilde{X}_3$ is the fully resolved elliptic Calabi-Yau threefold, the number of charged hypermultiplets is $H_{\rm charged}=58$, which is exactly the contribution from two copies of rank-1 E-string theory~\cite{Ganor:1996mu,Hanany:1997gh}.

Alternatively, one can choose to blow up the base $B_2$ at the two codimension-two (4,6) loci while keeping $H_{\rm neutral}$ unchanged, see figure~\ref{fig:12-E8}, leading to $T\rightarrow T+2=3$ and $H_{\rm charged}\rightarrow 0$, and (\ref{grav-anomaly}) still holds.

Now there is the third physical description of a single $(-10)$-curve, as shown in figure~\ref{fig:E7half}. Without the need to blow up the base $B_2$, we propose that there is a pure $\Ehalf$ gauge theory on the $(-10)$-curve. Now with $V=190$, $H_{\rm charged}=0$, $T=1$ and $H_{\rm neutral}=434$, the gravitational anomaly cancellation (\ref{grav-anomaly}) holds as well.

We now check the other anomaly cancellation conditions~\cite{Sagnotti:1992qw,Kumar:2010ru,Johnson:2016qar}:
\be
\ba
\label{ABC-anomaly}
-K\cdot \Sigma &=\frac{1}{6}\lambda (\sum x_R A_R-A_{adj.})\cr
\Sigma\cdot \Sigma &=\frac{1}{3}\lambda^2(\sum x_R C_R-C_{adj.})\cr
0&=\sum x_R B_R-B_{adj.}\,,
\ea
\ee
where $-K$ is the anticanonical class of the base $B_2$, $x_R$ is the multiplicity of matter in the representation $R$ of the gauge Lie algebra on the curve $\Sigma$. For $\Ehalf$, the coefficient $\lambda=12$, and the relevant $A_R$, $B_R$, $C_R$ coefficients are listed in table~\ref{t:ABC}, which we present the detailed derivation in Appendix~\ref{app:anomaly} following \cite{Kumar:2010ru}.

\begin{table}
\centering
\begin{tabular}{|c|c|c|c|}
\hline
\hline
$R$ & $A_R$ & $B_R$ & $C_R$\\
\hline
\hline
$\mbf{57}$ & 1 & 0 & $\frac{1}{24}$\\
\hline
$\mbf{190}$ & 4 & 0 & $\frac{5}{24}$\\
\hline
\end{tabular}
\caption{The $A_R$, $B_R$, $C_R$ coefficients for the representations $R=\mbf{57}$, $\mbf{190}$ of the $\Ehalf$ Lie algebra.}\label{t:ABC}
\end{table}

For a pure $\Ehalf$ gauge theory on a genus-0 $(-10)$-curve, the first and second equations in (\ref{ABC-anomaly}) can be verified, with the Riemann-Roch theorem $K\cdot \Sigma+\Sigma\cdot \Sigma=2g-2$:
\be
\ba
-K\cdot \Sigma&=-8=\frac{1}{6}\cdot 12\cdot (-4)\cr
\Sigma\cdot \Sigma&=-10=\frac{1}{3}\cdot (12)^2\cdot \left(-\frac{5}{24}\right)\,.
\ea
\ee
For $\Ehalf$ all the $B_R$-coefficients vanish, hence the third equation in (\ref{ABC-anomaly}) trivially holds. Thus, the pure $\Ehalf$ gauge theory over a $(-10)$-curve is completely anomaly-free.

\textit{5d prepotential for $\Ehalf$ gauge theory.} For additional physical evidence of the $\Ehalf$ gauge theory, we look at the dual 5d M-theory picture, which is obtained from circle compactification of the 6d theory. In this picture, the low energy dynamics of the 5d $\mc{N}=1$ theories is described by the prepotential in its Coulomb branch, where  the real scalar in the vector multiplets has non-trivial VEV $\phi$. The gauge algebra in the Coulomb branch is broken to $U(1)^{7}\subset E_7\subset \Ehalf$, so we can use the subalgebra $E_7$ to define the roots of $\Ehalf$. According to \cite{Intriligator:1997pq}, 
the prepotential for the low energy theory is
\be\label{eq:5dprepotential_KK}
\mathcal{F}=\frac{1}{12}\sum_{\alpha\in \Delta}|\alpha\cdot\phi|^3+\frac{1}{\gYM^2} \frac{1}{4h^{\vee}_G}\sum_{\alpha\in \Delta}(\alpha\cdot\phi)^2+\text{KK}\,,
\ee
where the set of roots $\Delta$ is consistent with weights in the ${\bf 133}+ {\bf 56}+ {\bf 1} $ representation of $E_7$. $h^{\vee}_G=24$ is the dual Coxeter number for $\Ehalf$. $\gYM$ is the 5d Yang-Mills coupling constant. $\text{KK}$ denotes the Kaluza-Klein contributions from the circle compactification of the 6d $\Ehalf$ theory (defined on a $(-10)$-curve as before). A full expression for the prepotential including the KK part will be presented in Appendix~\ref{app:prepotential}.

To verify the $\Ehalf$ prepotential from the Calabi-Yau threefold geometry, we put M-theory on a partially resolved $\overline{X}_3$ of the singular elliptic fibration $X_3$. The triple intersection numbers among the compact divisors exactly match the expected prepotential (\ref{eq:5dprepotential_KK}) for a pure $\Ehalf$ gauge theory. See Appendix~\ref{app:resol} for the details of the resolution sequence and triple intersection numbers for $\overline{X}_3$. 

\textit{Evidence from $\Ehalf$ elliptic genera.} The final concrete evidence is from the elliptic genera (EG) in the 6d (1,0) theory. Namely, we consider self-dual BPS strings  engineered in F-theory as D3-branes wrapping the $(-n)$-curve $\Sigma\sim S^2$.  For the 6d pure gauge theory on the  background $\mathbb{R}^4\times T^2$, the worldvolume theory for the D3-branes is a 4d $\mathcal{N}=2$ theory on $S^2\times T^2$ \cite{DelZotto:2016pvm,Beem:2019snk,Kapustin:2006hi,Putrov:2015jpa}. Compactifying the $S^2$ direction gives rise to a 2d $\mathcal{N}=(0,4)$ CFT on $T^2$, which is the worldsheet theory for the self-dual strings. It was proposed in \cite{DelZotto:2016pvm} that such a physical relation connects the elliptic genera for the self-dual string CFTs and the Schur indices for certain 4d $\mathcal{N}=2$ theories. This relation has been explicitly checked for $n=3,4,5,6,8,12$ where the 6d gauge symmetries (or the 4d flavor symmetries) are $SU(3),SO(8),F_4,E_6,E_7$ and $E_8$ respectively. In this letter, we generalize this correspondence by calculating the one-string elliptic genus to the  $n=10$ theory with the intermediate gauge Lie algebra $\Ehalf$. Our calculation has a perfect agreement with the generalized Schur index that has been recently proposed in \cite{Deb:2025ypl}.

On the tensor branch where a non-trivial expectation value $\phi_0$ for the scalar in the tensor multiplet is turned on, one can define the BPS partition function of the 6d theory that has the expression
\begin{align}
    Z_{\text{BPS}}=Z_{\text{pert}}\left(1+\sum_{k=1}^{\infty}e^{{k}\, \phi_0}\,\mathbb{E}_{{k}}(\tau;\epsilon_+,\epsilon_-)\right)\,.
\end{align}
where the $Z_{\text{pert}}$ factor includes contributions of towers of BPS particles from 6d supermultiplets with KK momentum $\tau$ along the 6d circle and $\mathbb{E}_{{k}}$ is the contribution from $k$ BPS strings wrapped around the 6d circle, which is the Ramond-Ramond (RR) elliptic genus for the 2d $\mathcal{N}=(0,4)$ worldsheet SCFT of $k$ self-dual strings on $T^2$.
The elliptic genus for the 2d theory is a twisted Witten index, by turning on fugacities for the global symmetries:
\be
SU(2)_-\times SU(2)_+\times SU(2)_{\mathcal{R}}\times G
\ee
where $\epsilon_{\pm}=\frac{1}{2}(\epsilon_1\pm\epsilon_2)$ are the fugacities of the $ SU(2)_+\times SU(2)_-$. $SU(2)_{\mathcal{R}}$ is the R-symmetry for the 2d theory. It is twisted with the $SU(2)_+$  so it has the same fugacities $\epsilon_+$. $G$ is the 6d gauge symmetry. $\tau$ is the complex structure of the $T^2$ that is identified with the axio-dilaton field in IIB. 

Under the modular transformation $S:\tau\rightarrow -1/\tau$, the elliptic genus transforms as
\begin{align}\label{eq:EkS_tran}
    \mathbb{E}_{{k}}(-1/\tau;-\epsilon_+/\tau,-\epsilon_-/\tau)=e^{\frac{2\pi i}{\tau}\mathcal{I}_{\mathrm{2d},{k}}}\mathbb{E}_{{k}}(\tau;\epsilon_+,\epsilon_-)\,,
\end{align}
where the modular anomaly index polynomial $\mathcal{I}_{2d,{k}}$ is captured by the 't Hooft anomaly for the 2d theory, which has been determined in \cite{Kim:2016foj,Shimizu:2016lbw} from the anomaly inflow as
\be
    \mathcal{I}_{\mathrm{2d},{k}}=-\epsilon_{+}^2(2-n+h_G^{\vee})k+\frac{1}{2}\epsilon_1\epsilon_2(nk^2+(2-n)k)\,,
\ee
where $h^{\vee}_G$ is the dual Coxeter number for the gauge algebra. In general, a holomorphic Jacobi form is a function $f_{k,m}(\tau;z)$ that has the transformation rule
\begin{align}
    f_{k,m}(-1/\tau;-z/\tau)=(-\tau)^{k}e^{-2\pi i\, mz^2/\tau}f_{k,m}(\tau;z)\,.
\end{align}
The numbers $k$ and $m$ are the weight and the index of the Jacobi form $f_{k,m}(\tau;z)$. 
The modular anomaly \eqref{eq:EkS_tran} suggests that the $k$-string elliptic genus is a weight zero meromorphic Jacobi form with index polynomial $\mathcal{I}_{\mathrm{2d},{k}}$, so that it can be bootstrapped via the modular bootstrap method proposed in \cite{DelZotto:2016pvm}. With proper modular ansatz and boundary conditions, we successfully obtain the exact expression for the one-string elliptic genus $\mathbb{E}_{1}$ in the 6d (1,0) pure $\Ehalf$ theory,  see detailed results in Appendix~\ref{app:EG}.

By factoring out the center of mass contributions, the interacting piece of the elliptic genus $\widetilde{\mathbb{E}}_{1}(v,q)={\mathbb{E}}_{1}(\tau;\epsilon_+,\epsilon_-)\theta_1(\epsilon_1)\theta_1(\epsilon_2)/\eta(\tau)^2$ has a $v=e^{2\pi i \epsilon_+}, q=e^{2\pi i \tau}$ series expansion \footnote{The overall factor is a consequence of the spectral flow symmetry as we will see later.}
\be\label{eq:5d_1inst}
    \widetilde{\mathbb{E}}_{1}(v,q)=q^{-\frac{h_G^{\vee}-1}{6}}v^{-\frac{h_G^{\vee}-n}{2}}\sum_{i,j\ge 0}(q/v^2)^iv^{2j}b_{ij}\,,
\ee
which exhibits many physical properties for the theory.

In the limit $q\rightarrow 0$, only the KK zero modes from circle compactification contribute. The one-string elliptic genus should be reduced to the one-instanton partition function for the 5d $\Ehalf$ pure gauge theory. This suggests that a boundary condition which we have used to fix the elliptic genus:
\be\ba
    \lim_{q\rightarrow0}&q^{\frac{23}{6}}\widetilde{\mathbb{E}}_{1}(v,q)=v^{h_G^{\vee}-1}\sum_{l=1}^{\infty}\dim(R_{l\theta_G}) v^{2l}\\
    &=v^{23}(1+190v^2+15504 v^4+749360 v^6+\cdots)\,,
\ea\ee
where $\dim(R_{l\theta_G})$ is the dimension for the representation whose highest weight is $l$ times the highest weight $\theta_G$ of the adjoint representation of $G=\Ehalf$~\cite{Lee:2023owa,Benvenuti:2010pq,Keller:2011ek}. 

The second property is the \emph{spectral flow symmetry} as suggested in \cite{DelZotto:2018tcj,DelZotto:2016pvm} that connects the Neveu-Schwarz-Ramond (NS-R) elliptic genus $\widetilde{\mathcal{E}}_1(v,q)$ and the R-R elliptic genus $\widetilde{\mathbb{E}}_1(v,q)$ via
\be
    \widetilde{\mathcal{E}}_1(v,q)=q^{\frac{n-h_G^{\vee}}4}v^{-(n-h_G^{\vee})}\widetilde{\mathbb{E}}_1(q^{1/2}/v,q)\,,
\ee
where $n-h_G^{\vee}=-14$ for $\Ehalf$.
The NS-R and R-R elliptic genera have different periodic conditions for left-moving fermions. In the 6d minimal pure gauge theory, the left-moving fermions are absent, indicating that these two types of elliptic genera are identical and providing the spectral flow symmetry up to an overall sign
\be\label{eq:SFsym}
    \widetilde{\mathbb{E}}_1(v,q)=(-1)^{n+1}q^{\frac{n-h_G^{\vee}}4}v^{-(n-h_G^{\vee})}\widetilde{\mathbb{E}}_1(q^{1/2}/v,q)\,,
\ee
for all the 6d pure gauge theories. Substituting the expansion \eqref{eq:5d_1inst} into \eqref{eq:SFsym}, we have for $G=\Ehalf$,
\be
 b_{i,j}=-b_{j,i}, \quad \forall i,j\geq 0\,.
\ee
which is indeed correct for our result.

Furthermore, as we have mentioned from the beginning of this subsection, the elliptic genus we have calculated is obtained from the worldvolume theory of a D3 on $\Sigma \times T^2$ and which is observed in \cite{DelZotto:2016pvm} that it is related to the Schur index of $S^3\times S^1$ by extracting the function $L_G(v,q)$, which corresponds to the flavored elliptic genus discussed in \cite{Putrov:2015jpa}. This observation will be reviewed in Appendix~\ref{app:EG} for self-consistency. For the special value $v=q^{1/4}$,
\be\ba\label{eq:LG}
L_G(q^{1/4},q)=1&+190 q^{1/2}+15695 q+783010 q^{3/2}\\
&+27319455 q^2+725679750 q^{5/2} +\mathcal{O}(q^3)\,,
\ea\ee
which is exactly the generalized Schur index proposed in \cite{Deb:2025ypl} for $\alpha=4$ and the level $-5$ character for the $\Ehalf$ VOA in \cite{Lee:2023owa}.

All these consistent checks for the elliptic genus provide strong supports on the existence for this 6d $\Ehalf$ SCFT.

\textit{6d $\Ehalf$ gauge theory with matter.} Besides the pure 6d (1,0) $\Ehalf$ gauge theory on a $(-10)$-curve, we now consider $\Ehalf$ on genus-$g$ curve $\Sigma$ with intersection number $-n$. We solve matter multiplicities $x_{\mbf{57}}$ and $x_{\mbf{190}}$ for the fundamental matter $\mbf{57}$ and the adjoint matter $\mbf{190}$ on $\Sigma$, using the anomaly cancellation equations (\ref{ABC-anomaly}) and coefficients in table~\ref{t:ABC}:
\be
\ba
-(2g-2+n)&=2\cdot(x_{\mbf{57}}+4x_{\mbf{190}}-4)\cr
-n&=48\cdot (\frac{1}{24}x_{\mbf{57}}+\frac{5}{24}x_{\mbf{190}}-\frac{5}{24})\,.
\ea
\ee
\be
\label{ng-x}
x_{\mbf{57}}=\frac{1}{2}(10-n-10g)\ ,\ x_{\mbf{190}}=g\,.
\ee

The first example is the $\Ehalf$ gauge theory on a genus-0, $(-9)$-curve, where (\ref{ng-x}) shows that there should be a half-hypermultiplet of the fundamental representation $\mbf{57}$. To understand this, we first decompose $\mbf{57}\rightarrow\mbf{56}+\mbf{1}$ in terms of the $E_7$ representations. The reality condition is only imposed on the pseudoreal part $\mbf{56}$, leading to $\frac{1}{2}\mbf{56}$. Effectively, in 6d (1,0) theory its contribution to $H_{\rm charged}$ is the same as 28 hypermultiplets, as the $\mbf{1}$ part is uncharged under the Cartan subalgebra of $\Ehalf$. For only one copy of half-hypermultiplet, it cannot be assigned a non-zero mass, and hence it cannot trigger the Higgs mechanism from $\Ehalf$ to $E_7$, which means that $\Ehalf+\frac{1}{2}\mbf{56}+\mbf{1}$ on a $(-9)$-curve can also be interpreted as a non-Higgsable cluster~\cite{Morrison:2012np}.

From the gravitational anomaly cancellation (\ref{grav-anomaly}) with $T=1$, $V=190$, $H_{\rm charged}=28$ in this case, we compute $H_{\rm neutral}=406$\footnote{This $H_{\rm neutral}$ differs from the case of $E_8$ with three copies of rank-1 E-string by 1, as $H_{\rm neutral}=405$ in the latter case.}.

The second example is the $\Ehalf$ gauge theory on a genus-0, $(-8)$-curve, with one fundamental matter $\mbf{57}$ from (\ref{ng-x}). Only the $\mbf{56}$ part of $\mbf{57}\rightarrow\mbf{56}+\mbf{1}$ is charged, hence the gravitational anomaly cancellation (\ref{grav-anomaly}) with $T=1$, $V=190$, $H_{\rm charged}=56$ leads to $H_{\rm neutral}=378$. As explained around (\ref{matter-Lag}), giving VEV to the scalars $\phi_{\mbf{56}}$ triggers a Higgs mechanism, after which $56$ out of the 190 vector multiplets become massive, and the gauge symmetry becomes $E_7\times U(1)$. After further decoupling the $U(1)$, we arrive at the non-Higgsable $E_7$ Yang-Mills theory on the $(-8)$-curve. 

As in the case of a $(-10)$-curve, after the same partial resolution sequence, the resulting 5d $\mc{N}=1$ prepotentials in the M-theory picture all match with the gauge theory result, see Appendix~\ref{app:resol}.

Finally, if $n=0$, $g=1$, i.e. putting the $\Ehalf$ gauge theory on a genus-1 curve $\Sigma$ with $\Sigma^2=0$, from (\ref{ng-x}) we get an adjoint matter $\mbf{190}$, similar to the cases of other gauge algebra~\cite{Katz:1996xe,Johnson:2016qar}.

\textit{(non-)Lagrangian interpretation of $\Ehalf$ gauge theory.} Given the Lie bracket and bilinear symmetric form $g_{ab}$ of $\Ehalf$, we can define the gauge field $\t{A}_\mu$, the covariant derivative $\t D_\mu=\ptl_\mu+\t A_\mu$, and the field strength $\t F_{\mu\nu}=\ptl_\mu \t A_\nu-\ptl_\nu \t A_\mu+[\t A_\mu,\t A_\nu]$ taking value in $\Ehalf$ in the usual way. The action of pure $\Ehalf$ Yang-Mills theory in $d$-spacetime dimensions is hence naturally defined as
\be
S=\int d^d x -\frac{1}{2\gYM^2}\tr(\t F_{\mu\nu}\t F^{\mu\nu})\,.
\ee
After we decompose the gauge fields $\t A_\mu\in\mbf{190}$ into $A_\mu\in\mbf{133}$, $B_\mu\in\mbf{56}$, $C_\mu\in\mbf{1}$, the action can be expanded as
\be
\label{E7-action}
S=\int d^d x -\frac{1}{2\gYM^2}\tr(F_{\mu\nu}F^{\mu\nu})\,,
\ee
where $F_{\mu\nu}=\ptl_\mu A_\nu-\ptl_\nu A_\mu+[A_{\mu},A_{\nu}]$ is the field strength for the $E_7$ gauge field $A_\mu$. Due to the degeneration of the bilinear symmetric form $g_{ab}$, the gauge fields $B_\mu$ and $C_\mu$ do not appear in the action~\ref{E7-action}. In other words, the $\Ehalf$ gauge theory should be interpreted as non-Lagrangian theory. For example, the aforementioned 6d (1,0) $\Ehalf$ gauge theory on a $(-10)$-curve has a UV completion as the 6d (1,0) SCFT in the zero volume limit of the $(-10)$-curve.

Moreover, from the decomposition (\ref{E7U1-branching}), the $\Ehalf\subset E_8$ subalgebra is not a real Lie subalgebra. Similar to the Heisenberg algebra, there does not exist a compact real Lie group corresponding to the real $\Ehalf$ Lie algebra. Nonetheless the computations for 6d and 5d $\Ehalf$ gauge theories are all performed at the level of the $E_7\subset \Ehalf$ subalgebra or the Cartan subalgebra $U(1)^7\subset E_7$, and they are not affected by these subtleties.

For the $\Ehalf$  matter fields, for example the action for $\Ehalf$-gauge theory coupled to a scalar field in the $\mbf{57}$ representation is
\be
S=\int d^d x -\frac{1}{2\gYM^2}\tr(\t F_{\mu\nu}\t F^{\mu\nu})+(\t D_\mu\phi)^\dagger(\t D^\mu\phi)\,.
\ee
After the decomposition of the matter fields $\mbf{57}$ into $\phi_{\mbf{56}}\in\mbf{56}$ and $\phi_{\mbf{1}}\in\mbf{1}$, from (\ref{57-action}) the action reads
\be
\ba
\label{matter-Lag}
S&=\int d^d x-\frac{1}{2\gYM^2}\tr( F_{\mu\nu} F^{\mu\nu})+\tr|\ptl_\mu\phi_{\mbf{56}}+A_\mu\cdot\phi_{\mbf{56}}|^2\cr
&+\frac{1}{2}|\ptl_\mu\phi_{\mbf{1}}+\omega(B_\mu,\phi_{\mbf{56}})|^2
\ea
\ee
where $\omega$ is the symplectic form in (\ref{omega}). After the scalars $\phi_{\mbf{56}}$ acquires a non-zero vev $\langle \phi_{\mbf{56}}\rangle\neq 0$, the gauge bosons acquire mass through Higgs mechanism. More precisely, the gauge field $C_\mu$ cannot acquire mass as it does not couple to $\phi_{\mbf{56}}$, and only 56 W-bosons can be Higgsed.

\vspace{0.3cm}
\begin{acknowledgments}

We would like to thank Kimyeong Lee for useful discussions. XW is supported by the Fundamental Research Funds for the Central Universities Grants No. WK2030250140 and the National Natural Science Foundation of China Grant No.12247103. YNW is supported by National Natural Science Foundation of China under Grant No. 12422503, No. 12247103.

\end{acknowledgments}


\bibliography{F-ref.bib}

\onecolumngrid

\appendix

\section{Anomaly coefficients for $\Ehalf$}
\label{app:anomaly}

We compute the coefficients $\lambda,A_R,B_R,C_R$ in the anomaly cancellation of 6d (1,0) $\Ehalf$ gauge theory, following \cite{Kumar:2010ru}. In general, from the coefficients for a subalgebra $H\subset G$, we can compute the coefficient $\lambda_G$ with
\be
\label{lambdaGH}
\lambda_G=\lambda_H\sum_i n(f)_i A_{S_i}(H)\,,
\ee
if the fundamental representation $f$ of $G$ has branching rule $f\rightarrow \sum n(f)_i S_i$ under $H\subset G$, where $S_i$ are irreducible representations of $H$.

For $A_R(G)$ of a representation $R$ of $G$, we have the formula
\be
\label{AGH}
A_R(G)=\frac{\sum_i n(R)_i A_{T_i}(H)}{\sum_i n(f)_i A_{S_i}(H)}
\ee
if $R$ has the branching rule $R\rightarrow \sum n(R)_i T_i$ under $H\subset G$.

For $C_R(G)$ of a representation $R$ of $G$, we have a similar formula
\be
\label{CGH}
C_R(G)=\frac{\sum_i n(R)_i C_{T_i}(H)}{(\sum_i n(f)_i A_{S_i}(H))^2}\,.
\ee

To compute the coefficients for $\Ehalf$ we use the following branching rules $E_7\subset \Ehalf$, where $H=E_7$, $G=\Ehalf$: 
\be
\label{190-branch}
\mbf{190}\rightarrow\mbf{133}+\mbf{56}+\mbf{1}\,,
\ee
\be
\label{57-branch}
\mbf{57}\rightarrow\mbf{56}+\mbf{1}\,,
\ee
and the relevant coefficients for $E_7$~\cite{Johnson:2016qar}:
\be
\lambda_{E_7}=12\ ,\ A_{\mbf{56}}(E_7)=1\ ,\ A_{\mbf{133}}(E_7)=3\ ,\ C_{\mbf{56}}(E_7)=\frac{1}{24}\ ,\ C_{\mbf{133}}(E_7)=\frac{1}{6}\,.
\ee
First, from (\ref{lambdaGH}) and (\ref{57-branch}) we can compute the $\lambda_{\Ehalf}$ as
\be
\ba
\lambda_{\Ehalf}&=\lambda_{E_7}\times A_{\mbf{56}}\cr
&=12\,.
\ea
\ee
From (\ref{AGH}) we get
\be
A_{\mbf{57}}=1\ ,\ A_{\mbf{190}}=4\,.
\ee
From (\ref{CGH}) we get
\be
C_{\mbf{57}}=\frac{1}{24}\ ,\ C_{\mbf{190}}=\frac{5}{24}\,.
\ee
The $B_R$ coefficients of $\Ehalf$ all vanish, similar to the other exceptional Lie algebras.

\section{Partial resolution of the Weierstrass model on $\mb{F}_{10}$}
\label{app:resol}

We provide further evidence for the pure $\Ehalf$ theory, from the partial resolution of the singular Weierstrass model $X_3$ over $\mb{F}_{10}$. We define the compact, singular elliptic threefold $X_3$ as an anticanonical hypersurface of a toric ambient space $Y_4$, that is the blow-up of a $\mb{P}^{2,3,1}$ fibration over $\mb{F}_{10}$. We start from the 4d polytope with vertices
\be
\ba
\label{F10-vertices}
&x:(0,0,1,0)\ ,\ y:(0,0,0,1)\ ,\ z:(0,0,-2,-3)\ ,\ u:(0,-1,-2,-3)\ ,\ v:(1,0,-2,-3)\ ,\ \cr
&s:(-1,-10,-2,-3)\ ,\ t:(0,1,-2,-3)\,,
\ea
\ee
and perform the blow-up sequence of $Y_4$ corresponding to an $E_7$ singularity in F-theory, following the notations of \cite{Lawrie:2012gg}:
\be
\ba
\label{E7half-resolution}
&(x,y,u;u_1),(x,y,u_1;u_2),(y,u_1;u_3),(y,u_2;u_4),(u_2,u_3;u_5),(u_1,u_3;u_6),\cr
&(u_2,u_4;u_7),(u_3,u_4;u_8),(u_4,u_5;u_9),(u_5,u_8;u_{10}),(u_3,u_5;u_{11})\,.
\ea
\ee
For instance $(x,y,u;u_1)$ means the replacement $x\rightarrow xu_1$, $y\rightarrow yu_1$, $u\rightarrow uu_1$ and the new $[x:y:u]$ are projective coordinates satisfying $(x,y,u)\sim (\lambda x,\lambda y,\lambda u)$ for $\lambda\in\mb{C}^*$.

The new exceptional divisors of the blown up 4d polytope $\overline{Y}_4$ correspond to the rays:
\be
\ba
\label{Exceptional-rays}
&u_1:(0,-1,-1,-2)\ ,\ u_2:(0,-1,0,-1)\ ,\ u_3:(0,-1,-1,-1)\ ,\ u_4:(0,-1,0,0)\cr
&u_5:(0,-2,-1,-2)\ ,\ u_6:(0,-2,-2,-3)\ ,\ u_7:(0,-2,0,-1)\ ,\ u_8:(0,-2,-1,-1)\cr
&u_9:(0,-3,-1,-2)\ ,\ u_{10}:(0,-4,-2,-3)\ ,\ u_{11}:(0,-3,-2,-3)\,.
\ea
\ee

The anticanonical CY3 in $\overline{Y}_4$ has been partially resolved to $\overline{X}_3$. The effective exceptional divisors are $D_1:u_6=0$, $D_2:u_{11}=0$, $D_3:u_{10}=0$, $D_4:u_9=0$, $D_5:u_7=0$, $D_6:u_4=0$, $D_7:u_8=0$. We also denote $D_0:u=0$ as the vertical divisor corresponding to the affine node over the $(-10)$-curve.

We can compute the triple intersection number between these $D_i$s from standard toric geometry methods, e.g. as in \cite{Apruzzi:2019opn,Apruzzi:2019kgb,Closset:2020scj}. Despite of the singularity of $\overline{X}_3$, all the triple intersection numbers are integral (we only list the non-zero ones):
\be
\ba
\label{10-Int}
&D_4\cdot D_5\cdot D_6=-2\ ,\ D_4\cdot D_6\cdot D_7=-2\ ,\ D_3\cdot D_4\cdot D_7=-2\ ,\ D_6\cdot D_5^2=-2\ ,\ D_6^2\cdot D_7=2\ ,\ \cr
&D_6\cdot D_7^2=2\ ,\ D_6^2\cdot D_4=2\ ,\ D_6\cdot D_4^2=2\ ,\ D_5\cdot D_4^2=-4\ ,\ D_5^2\cdot D_4=2\ ,\ D_3\cdot D_2^2=2\ ,\ \cr
&D_3^2\cdot D_2=-4\ ,\ D_3\cdot D_7^2=-2\ ,\ D_7\cdot D_4^2=2\ ,\ D_7^2\cdot D_4=2\ ,\ D_4^2\cdot D_3=-2\ ,\ D_1\cdot D_2^2=-6\ ,\ \cr
&D_1^2\cdot D_2=4\ ,\ D_0\cdot D_1^2=-8\ ,\ D_1\cdot D_0^2=6\ ,\ \cr
&D_0^3=8\ ,\ D_1^3=8\ ,\ D_2^3=8\ ,\ D_3^3=8\ ,\ D_4^3=10\ ,\ D_5^3=8\ ,\ D_6^3=10\ ,\ D_7^3=10\,.
\ea
\ee

The Dynkin diagram structure of $\Ehalf$ can be read off from the intersection numbers with another vertical divisor $D:v=0$:
\be
\ba
&D\cdot D_i^2=-2\ (i=0,...,7)\ ,\ D\cdot D_0\cdot D_1=D\cdot D_1\cdot D_2=D\cdot D_2\cdot D_3=D\cdot D_3\cdot D_4=1\cr
&D\cdot D_4\cdot D_5=D\cdot D_5\cdot D_6=D\cdot D_3\cdot D_7=1\,.
\ea
\ee
One can see from $D_4^3=D_6^3=D_7^3=10$, these abnormal divisors indeed correspond to the fermionic nodes of the Dynkin diagram of $\Ehalf$~\cite{Lee:2023owa}.

Similar to the cases of other non-Higgsable clusters, we can decompactify all divisors except for $D_1\sim D_7$, to get a non-compact CY3 potentially leading to 5d pure $\Ehalf$ gauge theory~\cite{Jefferson:2017ahm,Bhardwaj:2018yhy,Apruzzi:2019enx}. This involves decoupling the 5d vector multiplet corresponding to $D_0$ as well~\cite{Bhardwaj:2019xeg,Apruzzi:2019kgb}. The cubic terms of the prepotential of this pure $\Ehalf$ gauge theory can be read off from the triple intersection numbers:
\be\label{eq:Fcubic}
\ba
\mc{F}_{\rm cubic}&=\frac{1}{6}\sum_{i,j,k=1}^7(D_i\cdot D_j\cdot D_k)\phi_i\phi_j\phi_k\cr
&=\frac{1}{6}(8\phi_1^3+8\phi_2^3+8\phi_3^3+10\phi_4^3+8\phi_5^3+10\phi_6^3+10\phi_7^3+12\phi_1^2\phi_2-18\phi_1\phi_2^2\cr
&+6\phi_2^2\phi_3-12\phi_2\phi_3^2-6\phi_3\phi_4^2+6\phi_4\phi_5^2-12\phi_4^2\phi_5+6\phi_4^2\phi_6+6\phi_4\phi_6^2\cr
&+6\phi_6\phi_7^2+6\phi_6^2\phi_7-6\phi_5^2\phi_6-6\phi_3\phi_7^2+6\phi_4^2\phi_7+6\phi_4\phi_7^2\cr
&-12\phi_3\phi_4\phi_7-12\phi_4\phi_5\phi_6-12\phi_4\phi_6\phi_7)\,.
\ea
\ee

For the 6d KK theory, we add in the vertical divisor $D_0$ (corresponding to $\phi_0$), to get the prepotential from triple intersection numbers:
\be
\ba
\label{10-prepotential}
\mc{F}_{\rm cubic}
&=\frac{1}{6}(8\phi_0^3+8\phi_1^3+8\phi_2^3+8\phi_3^3+10\phi_4^3+8\phi_5^3+10\phi_6^3+10\phi_7^3+12\phi_1^2\phi_2-18\phi_1\phi_2^2\cr
&+6\phi_2^2\phi_3-12\phi_2\phi_3^2-6\phi_3\phi_4^2+6\phi_4\phi_5^2-12\phi_4^2\phi_5+6\phi_4^2\phi_6+6\phi_4\phi_6^2\cr
&+6\phi_6\phi_7^2+6\phi_6^2\phi_7-6\phi_5^2\phi_6-6\phi_3\phi_7^2+6\phi_4^2\phi_7+6\phi_4\phi_7^2\cr
&-12\phi_3\phi_4\phi_7-12\phi_4\phi_5\phi_6-12\phi_4\phi_6\phi_7-24\phi_0\phi_1^2+18\phi_1\phi_0^2)\,.
\ea
\ee

The triple intersection numbers (\ref{10-prepotential}) and prepotential (\ref{10-prepotential}) match that of $\Ehalf$ gauge theory with 190 vector multiplets, see Appendix~\ref{app:prepotential}.

After the partial crepant resolution (\ref{E7half-resolution}), we assign the redundant coordinates $u_1=u_2=u_3=u_5=1$, as well as $z=1$, and the lowest order terms in the resolved equation of $\overline{X}_4$ reads
\be
\ba
\label{E7half-resolved}
&u^5 u_{10}^2 u_{11}^3 u_6^4 u_8 u_9 (1 + av+bv^2) + 
 u^4 u_{10}^4 u_{11}^4 u_4  u_6^4 u_7^2 u_8^2 u_9^3 x+u_4 u_7^2 u_9 x^3 -u_4 u_8 y^2=0\,.
\ea
\ee
The projective relations among the coordinates are
\be
\ba
&[x u_4 u_7^2 u_8 u_9^2 u_{10}^2 u_{11}:y u_4^2 u_6 u_7^3 u_8^3 u_9^4 u_{10}^5 u_{11}^3:u]\cr
&[x:y u_4 u_6 u_7 u_8^2 u_9^2 u_{10}^3 u_{11}^2:u_6^2 u_8 u_9 u_{10}^2 u_{11}^2]\cr
&[y u_4 u_7 u_8 u_{10}:u_6]\cr
&[y:u_7 u_9 u_{10} u_{11}]\cr
&[u_7:u_6 u_8 u_{10} u_{11}]\cr
&[u_{11}:u_4 u_9]\cr
&[u_4:u_{10}u_{11}]\cr
&[u_{11}:u_8]\,.\cr
\ea
\ee

The equation (\ref{E7half-resolved}) is still singular at $u_4=u_8=u_9=0$, corresponding to the intersection locus $D_4\cdot D_6\cdot D_7$. The local equation near $u_4=u_8=u_9=0$ takes the form of (after assigning the other non-vanishing coordinates to 1)
\be
\label{cDV}
u_8 u_9(1+av+bv^2)+u_4 u_9-u_4 u_8=0\,,
\ee
which is a type $A_1$ surface singularity at $u_4=u_8=u_9=0$ fibered over the $v$ direction, with two points of further degeneration at the solutions of $1+av+bv^2=0$.

In fact, if we further resolve $(u_4,u_8,u_9;u_{12})$, we will reach the 5d description for an $E_8$ gauge theory coupled to two copies of rank-1 E-string.

We also present the triple intersection numbers for the partially resolved $\overline{X}_3$ if we replace the base $\mb{F}_{10}$ by $\mb{F}_9$, for the context of $\Ehalf+\frac{1}{2}\mbf{56}+\mbf{1}$. In the 5d M-theory description, we use the same partial resolution sequence (\ref{E7half-resolution}). Comparing to the setup before, the only difference is that we use the following set of vertices
\be
\ba
\label{F9-vertices}
&x:(0,0,1,0)\ ,\ y:(0,0,0,1)\ ,\ z:(0,0,-2,-3)\ ,\ u:(0,-1,-2,-3)\ ,\ v:(1,0,-2,-3)\ ,\ \cr
&s:(-1,-9,-2,-3)\ ,\ t:(0,1,-2,-3)\,.
\ea
\ee

After the partial resolution (\ref{E7half-resolution}), the non-zero triple intersection numbers between $D_1:u_6=0$, $D_2:u_{11}=0$, $D_3:u_{10}=0$, $D_4:u_9=0$, $D_5:u_7=0$, $D_6:u_4=0$, $D_7:u_8=0$, $D_0:u=0$ are
\be\label{9-Int}
\ba
&D_4\cdot D_5\cdot D_6=-1\ ,\ D_4\cdot D_6\cdot D_7=-1\ ,\ D_3\cdot D_4\cdot D_7=-1\ ,\ D_6^2\cdot D_5=2\ ,\ D_6\cdot D_5^2=-4\ ,\ \cr
&D_6^2\cdot D_7=1\ ,\ D_6\cdot D_7^2=1\ ,\ D_6^2\cdot D_4=1\ ,\ D_6\cdot D_4^2=1\ ,\ D_5\cdot D_4^2=-4\ ,\ D_5^2\cdot D_4=2\ ,\ \cr
&D_3\cdot D_2^2=1\ ,\ D_3^2\cdot D_2=-3\ ,\ D_3\cdot D_7^2=-1\ ,\ D_3^2\cdot D_7=-1\ ,\ D_7\cdot D_4^2=1\ ,\ D_7^2\cdot D_4=1\ ,\ \cr
&D_4^2\cdot D_3=-1\ ,\ D_4\cdot D_3^2=-1\ ,\ D_1\cdot D_2^2=-5\ ,\ D_1^2\cdot D_2=3\ ,\ D_0\cdot D_1^2=-7\ ,\ D_1\cdot D_0^2=5\ ,\ \cr
&D_0^3=8\ ,\ D_1^3=8\ ,\ D_2^3=8\ ,\ D_3^3=8\ ,\ D_4^3=9\ ,\ D_5^3=8\ ,\ D_6^3=9\ ,\ D_7^3=9\ ,\ \cr
&D\cdot D_i^2=-2\ (i=0,...,7)\ ,\ D\cdot D_0\cdot D_1=D\cdot D_1\cdot D_2=D\cdot D_2\cdot D_3=D\cdot D_3\cdot D_4=1\cr
&D\cdot D_4\cdot D_5=D\cdot D_5\cdot D_6=D\cdot D_3\cdot D_7=1\,.
\ea
\ee

Finally, for the case of $\Ehalf$ over a $(-8)$-curve, the triple intersection numbers are same as the generic elliptic fibration over the $(-8)$-curve, which we recall~\cite{DelZotto:2017pti}:
\be\label{8-Int}
\ba
&D_6^2\cdot D_5=4\ ,\ D_6\cdot D_5^2=-6\ ,\ D_5\cdot D_4^2=-4\ ,\ D_5^2\cdot D_4=2\ ,\ D_3^2\cdot D_2=-2\ ,\ D_3^2\cdot D_7=-2\ ,\ \cr
&D_4\cdot D_3^2=-2\ ,\ D_1\cdot D_2^2=-4\ ,\ D_1^2\cdot D_2=2\ ,\ D_0\cdot D_1^2=-6\ ,\ D_1\cdot D_0^2=4\ ,\ \cr
&D_0^3=8\ ,\ D_1^3=8\ ,\ D_2^3=8\ ,\ D_3^3=8\ ,\ D_4^3=8\ ,\ D_5^3=8\ ,\ D_6^3=8\ ,\ D_7^3=8\ ,\ \cr
&D\cdot D_i^2=-2\ (i=0,...,7)\ ,\ D\cdot D_0\cdot D_1=D\cdot D_1\cdot D_2=D\cdot D_2\cdot D_3=D\cdot D_3\cdot D_4=1\ ,\ \cr
&D\cdot D_4\cdot D_5=D\cdot D_5\cdot D_6=D\cdot D_3\cdot D_7=1\,.
\ea
\ee

\section{Prepotentials for 6d $\Ehalf$ theories}\label{app:prepotential}
Consider a 6d $(1,0)$ $\Ehalf$ supergravity theory engineered from F-theory compactification on an elliptically fibered Calabi-Yau threefold $X_3$ over $\mathbb{F}_n$ for $n\leq 10$. In the 5d KK theory obtained from the 6d theory on a circle, we turn on Coulomb parameters $\phi_i,i=1,\cdots,7,$ associated with the gauge algebra $U(1)^7\subset E_7 \subset \Ehalf$, together with tensor parameters $t_{b_i}$ and KK parameter $\tau$, the prepotential can be written as \cite[(A.33)]{Huang:2025xkc}
\be\label{eq:prepotential}
\mathcal{F}=\frac{1}{12}\sum_{\alpha\in \Delta} |\alpha\cdot \phi|^3-\frac{x_{\bf 57}}{6}\sum_{\omega\in R_{\mathbf{57}}} |\omega\cdot \phi|^3+\left(t_{b_2}-\frac{n-2}{2}\tau\right)\frac{1}{4h_G^{\vee}}\sum_{\alpha\in \Delta}(\alpha\cdot \phi)^2-\frac{1}{2}\Omega^{-1}_{ij}t_{b_i}t_{b_j}\tau-\frac{9-T}{24}\tau^3\,,
\ee
where $\Delta$ is the collection of weights in the representation $\mathbf{133}+\mathbf{56}+\mathbf{1}$ and $R_{\bf 57}$ is the set of  weights in the representation $\mathbf{56}+\mathbf{1}$. $x_{\bf 57}=\frac{1}{2}(10-n)$ is the number of hypermultiplets that transform under $\bf 57$ of $\Ehalf$. $T=1$ is the rank of the tensor multiplet. $t_{b_1}$ and $t_{b_2}$ correspond to the volumes of two $\mathbb{P}^1$'s in $\mathbb{F}_{10}$, and $\Omega_{ij}=\begin{pmatrix}
    0&1\\
    1&-n
\end{pmatrix}_{ij}$ is the intersection matrix between these two $\mathbb{P}^1$'s. Denote the first and second terms in \eqref{eq:prepotential} by $\mathcal{F}_{\text{vec}}$ and $x_{\bf 57}\mathcal{F}_{\text{hyp}}$ respectively. We find $\mathcal{F}_{\text{vec}}=\mathcal{F}_{\text{cubic}}$ defined in \eqref{eq:Fcubic} and 
\be\ba
6\,\mathcal{F}_{\text{hyp}}=&\,2\phi_4^3-6\phi_3\phi_4^2+6\phi_6\phi_4^2+6\phi_7\phi_4^2+6\phi_3^2\phi_4+6\phi_6^2\phi_4+6\phi_7^2\phi_4-12\phi_5\phi_6\phi_4-12\phi_3\phi_7\phi_4-12\phi_6\phi_7\phi_4+2\phi_6^3\\
&+2\phi_7^3-6\phi_1\phi_2^2-6\phi_2\phi_3^2-12\phi_5\phi_6^2-6\phi_3\phi_7^2+6\phi_6\phi_7^2+6\phi_1^2\phi_2+6\phi_2^2\phi_3+12\phi_5^2\phi_6+6\phi_3^2\phi_7+6\phi_6^2\phi_7\,,
\ea\ee
which agrees with the intersection numbers calculated in \eqref{10-Int}, \eqref{9-Int} and \eqref{8-Int} via
\begin{align}
    \mathcal{F}_{\text{vec}}+x_{\bf 57}\mathcal{F}_{\text{hyp}}=\frac{1}{6}\sum_{i,j,k=1}^{7}(D_i\cdot D_j\cdot D_k) \phi_i\phi_j\phi_k,
\end{align}
for $x_{\bf 57}=0,\frac{1}{2},$ and $1$ respectively.

The relevant tree level part of the prepotential also agree with the $D$ relevant intersection numbers via
\be\ba
    \frac{1}{4h_G^{\vee}}\sum_{\alpha\in \Delta}(\alpha\cdot \phi)^2=\,&\phi_1^2+\phi_2^2+\phi_3^2+\phi_4^2+\phi_5^2+\phi_6^2+\phi_7^2-\phi_1\phi_2-\phi_2\phi_3-\phi_3\phi_4-\phi_4\phi_5-\phi_5\phi_6-\phi_3\phi_7,\\
    =\,&\frac{1}{2}\sum_{i,j=1}^7 (D\cdot D_i\cdot D_j) \phi_i\phi_j\,.
\ea\ee

Finally, to extract the intersection numbers involving $D_0$, we need to shift each Coulomb parameters by comarks times $\phi_0$ \cite{Bhardwaj:2019fzv}. Since the parameters are expressed in the $E_7$ basis, the relevant comarks are those of $E_7$, which gives the shift
\begin{align}
    \phi_i\rightarrow \phi_i+b^{\vee}_i\phi_0, \qquad b^{\vee}=(2, 3, 4, 3, 2, 1, 2)\,.
\end{align}
We also shift the tensor parameters as
\begin{align}
    t_{b_1}\rightarrow t_{b_1}-\phi_0,\quad t_{b_2}\rightarrow t_{b_2}+n\phi_0.
\end{align}
We have verified that the $\phi_0$ dependent terms after the shift also agree with the $D_0$ relevant intersection numbers presented in \eqref{10-Int}, \eqref{9-Int} and \eqref{8-Int}.

In the 5d limit by decoupling all the KK modes from circle compactification, the combination $\left(t_{b_2}-\frac{n-2}{2}\tau\right)$ reduces to the 5d Yang-Mills coupling $1/g^2_{\text{YM}}$, recovering \eqref{eq:5dprepotential_KK}.

\section{One-string elliptic genus for 6d $\Ehalf$ SCFTs}\label{app:EG}

In this appendix, we provide details of the calculation of the elliptic genus for the 6d pure $\Ehalf$ theory. The method here was first proposed in \cite{DelZotto:2016pvm} for 6d pure gauge theories that are engineered from F-theory on elliptically fibered Calabi-Yau threefolds over $(-n)$-curves. For $n=3,4,5,6,8,12$, the corresponding gauge symmetries are $SU(3),SO(8),F_4,E_6,E_7$ and $E_8$ respectively. The corresponding dual Coxeter numbers are
\be
\begin{tabular}{|c|c|c|c|c|c|c|c|}
\hline
$G$ & $SU(3)$ & $SO(8)$ & $F_4$ & $E_6$ & $E_7$ & $\Ehalf$ & $E_8$\\
\hline
$n$ & $3$ & $4$ & $5$ & $6$ & $8$ & $10$ & $12$\\
\hline
$h_G^{\vee}$ & $3$ & $6$ & $9$ & $12$ & $18$ & $24$ & $30$\\
\hline
\end{tabular}
\ee
where $n,h_G^{\vee}$ satisfy $h_G^{\vee}=3n-6$ due to the anomaly cancellation. 
It was proposed in \cite{DelZotto:2016pvm} that the one-string elliptic genera for all these theories have a universal expression:
\begin{align}\label{eq:E1ansatz}
    \mathbb{E}_{1}(\tau;\epsilon_+,\epsilon_-)=\frac{\mathcal{N}(\tau,2\epsilon_+)}{\eta(\tau)^{4h_G^{\vee}-6}\,\theta_1(\epsilon_1)\theta_1(\epsilon_2)\, \varphi_{-2,1}(2\epsilon_+)^{h_G^{\vee}-1}}\,.
\end{align}
The functions $\theta_I(z)=\theta_I(\tau;z),I=1,2,3,4$ and $\eta(\tau)$ are the standard Jacobi theta functions and Dedekind eta function respectively. They are weight $1/2$ Jacobi forms that have indices $\frac{1}{2}$ and $0$ respectively. $\varphi_{k,m}(z)$ is a Jacobi form of weight $k$ and index $m$. For those we will be using in the letter, they are defined as
\be\ba
    \varphi_{0,1}(z)=4\sum_{I=2}^4\frac{\theta_I(z)^2}{\theta_I(0)^2},\qquad\varphi_{-2,1}(z)=\frac{\theta_1(z)^2}{\eta(\tau)^6}, \quad\text{and} \quad\varphi_{0,\frac{3}{2}}(z)=\frac{\theta_1(2z)}{\theta_1(z)}\,.
\ea\ee
We conjecture that the structure \eqref{eq:E1ansatz} still holds for the 6d pure $\Ehalf$ theory, where $h_G^{\vee}=24$.
Since the modular weight for the elliptic genus is zero and the Jacobi-form index is determined by the anomaly polynomial, we deduce that the numerator $\mathcal{N}(\tau,2\epsilon_+)$ is a Jacobi form of weight 0
and index $18+\frac{3}{2}$.
By using the generating rings for the Jacobi forms $\{E_4(\tau),E_6(\tau),\varphi_{-2,1}(z),\varphi_{0,1}(z)\}$, the numerator has only 37 undetermined coefficients $c_{l_i}$:
\be\ba\label{eq:6dE7halfansatz}
    \mathcal{N}&(\tau,2\epsilon_+)=\\
    &\quad\varphi_{0,\frac{3}{2}}(2\epsilon_+)\sum_{l_i \ge 0}c_{l_i}E_4^{l_1}E_6^{l_2}\varphi_{-2,1}(2\epsilon_+)^{l_3}\varphi_{0,1}(2\epsilon_+)^{l_4}\,.
\ea\ee
The coefficients $c_{l_i}$ can be determined from the boundary conditions of the elliptic genus.

The first boundary condition comes from the 5d limit of the 6d theory. Let $q=e^{2\pi i \tau}$, $v=e^{2\pi i \epsilon_+}$ and $x=e^{2\pi i \epsilon_-}$. In the limit $q\rightarrow 0$, the one-string elliptic genus reduces to the one-instanton partition function of the 5d pure $\Ehalf$ theory, as proposed in \cite{Lee:2023owa,Benvenuti:2010pq}
\be\label{eq:5d_1inst2}
    \lim_{q\rightarrow0}q^{4}{\mathbb{E}}_{1}(v,q)=\frac{v^{h_G^{\vee}}}{(1-vx)(1-v/x)}\sum_{l=1}^{\infty}\dim(R_{l\theta_G}) v^{2l}
    =\frac{v^{24}}{(1-vx)(1-v/x)}(1+190v^2+15504 v^4+749360 v^6+\cdots)\,.
\ee
Here the factor $q^4=q^{\frac{n-2}{2}}$ is the shift from the convention between the 6d tensor parameter $t_{b_2}$ and the 5d instanton counting parameter $1/g^2_{\text{YM}}=t_{b_2}-\frac{n-2}{2}\tau$. The dimension of the representation $R_{l\theta_G}$ has a universal formula \cite{Sextonions}
\begin{align}
    \dim(R_{l\theta_G})=\frac{3c+2l+5}{3c+5}\frac{\binom{l+2c+3}{l}\binom{l+5c/2+3}{l}\binom{l+3c+4}{l}}{\binom{l+c/2+1}{l}\binom{l+c+1}{l}},\quad c=\frac{1}{3}h_G^{\vee}-2\,.
\end{align}
The second condition is from the spectral flow symmetry which requires that the normalized elliptic genus has the expansion \eqref{eq:5d_1inst}. This indicates the vanishing coefficients for the first few terms in the $v$ expansion. These two conditions completely determines all the coefficients $c_{l_i}$. We find
\be\ba
\mathcal{N}(\tau,2\epsilon_+)=&\frac{\varphi_{0,\frac{3}{2}}}{2^{34}\times 3^{17}} \bigg(9504 \varphi_{-2,1}^{11} \varphi_{0,1}^7 E_6 E_4 \left(73262277 E_4^3-10938176 E_6^2\right)+114400 \varphi_{-2,1}^9 \varphi_{0,1}^9 E_6 \big(252800 E_6^2\\
&-8431263 E_4^3\big)-3359070 \varphi_{-2,1}^8 \varphi_{0,1}^{10} E_4 \left(201667 E_4^3-72384 E_6^2\right)+701220 \varphi_{-2,1}^6 \varphi_{0,1}^{12} \left(801567 E_4^3-60064 E_6^2\right)\\
&+652454356320 \varphi_{-2,1}^7 \varphi_{0,1}^{11} E_6 E_4^2-2592 \varphi_{-2,1}^{17} \varphi_{0,1} E_6 E_4 \left(-998080 E_6^2 E_4^3+87040 E_6^4+226611 E_4^6\right)\\
&+3360 \varphi_{-2,1}^{15} \varphi_{0,1}^3 E_6 \left(-11049408 E_6^2 E_4^3+164864 E_6^4+7724241 E_4^6\right)+540 \varphi_{-2,1}^{14} \varphi_{0,1}^4 E_4 (-250050528 E_6^2 E_4^3\\
&+24743936 E_6^4+33551253 E_4^6)-84 \varphi_{-2,1}^{12} \varphi_{0,1}^6 \left(-4894115904 E_6^2 E_4^3+91654144 E_6^4+1664997363 E_4^6\right)\\
&-6048 \varphi_{-2,1}^{13} \varphi_{0,1}^5 E_6 E_4^2 \left(37264779 E_4^3-18345152 E_6^2\right)+8910 \varphi_{-2,1}^{10} \varphi_{0,1}^8 E_4^2 \left(49226931 E_4^3-54938240 E_6^2\right)\\
&-27 \varphi_{-2,1}^{16} \varphi_{0,1}^2 E_4^2 \left(-478341504 E_6^2 E_4^3+172421120 E_6^4+22934097 E_4^6\right)-210281853600 \varphi_{-2,1}^5 \varphi_{0,1}^{13} E_6 E_4\\
&+25785477600 \varphi_{-2,1}^3 \varphi_{0,1}^{15} E_6-254970873900 \varphi_{-2,1}^4 \varphi_{0,1}^{14} E_4^2+59467757715 \varphi_{-2,1}^2 \varphi_{0,1}^{16} E_4\\
&+\varphi_{-2,1}^{18} \left(-133249536 E_6^2 E_4^6+169979904 E_6^4 E_4^3-2490368 E_6^6+1436859 E_4^9\right)-5571053555 \varphi_{0,1}^{18}\bigg)\,.
\ea\ee
By expanding the $\widetilde{\mathbb{E}}_{1}(v,q)$ with $q,v$ as we have done in \eqref{eq:5d_1inst}, we obtain the coefficients $b_{i,j}$ which are presented in \eqref{tab:bij}. 
\begin{table}[htbp]\label{tab:bij}
\centering
\scalebox{0.5}{
\begin{tabular}{|c|cccccccccccccccc|}
\hline
$i\backslash j$ & 0 & 1 & 2 & 3 & 4 & 5 & 6 & 7 & 8 & 9 & 10 & 11 & 12 & 13 & 14 & 15 \\
\hline
0  & 0 & 0 & 0 & 0 & 0 & 0 & 0 & 0 & 1 & 190 & 15504 & 749360 & 24732110 & 605537790 & 11619550320 & 181746027600 \\
1  & 0 & 0 & 0 & 0 & 0 & 0 & 0 & 0 & 0 & 191 & 33650 & 2568819 & 117126070 & 3673669570 & 86026550610 & 1587521062470 \\
2  & 0 & 0 & 0 & 0 & 0 & 0 & 0 & 0 & 0 & 0 & 18527 & 3015890 & 215308893 & 9270547450 & 276775252450 & 6209911975410 \\
3  & 0 & 0 & 0 & 0 & 0 & 0 & 0 & 0 & 0 & -1 & 0 & 1214768 & 182157600 & 12163627347 & 495205128390 & 14096056649296 \\
4  & 0 & 0 & 0 & 0 & 0 & 0 & 0 & 0 & 0 & 0 & -191 & 0 & 60439480 & 8332648800 & 520780842570 & 20076314098740 \\
5  & 0 & 0 & 0 & 0 & 0 & 0 & 0 & 0 & 0 & 0 & 0 & -18527 & 0 & 2428876217 & 307670159230 & 18017053206978 \\
6  & 0 & 0 & 0 & 0 & 0 & 0 & 0 & 0 & 0 & 0 & 0 & -190 & -1214768 & 0 & 81976263336 & 9545387953870 \\
7  & 0 & 0 & 0 & 0 & 0 & 0 & 0 & 0 & 0 & 0 & 0 & 0 & -33650 & -60439480 & 0 & 2386587310945 \\
8  & -1 & 0 & 0 & 0 & 0 & 0 & 0 & 0 & 0 & 2 & 0 & 0 & 0 & -3015890 & -2428876217 & 0 \\
9  & -190 & -191 & 0 & 1 & 0 & 0 & 0 & 0 & -2 & 0 & 382 & 380 & 0 & -15504 & -182157600 & -81976263336 \\
10 & -15504 & -33650 & -18527 & 0 & 191 & 0 & 0 & 0 & 0 & -382 & 0 & 37054 & 67300 & 31008 & -2568819 & -8332648800 \\
11 & -749360 & -2568819 & -3015890 & -1214768 & 0 & 18527 & 190 & 0 & 0 & -380 & -37054 & 0 & 2429536 & 6031780 & 5137638 & -213810173 \\
12 & -24732110 & -117126070 & -215308893 & -182157600 & -60439480 & 0 & 1214768 & 33650 & 0 & 0 & -67300 & -2429536 & 0 & 120878960 & 364315200 & 429868426 \\
13 & -605537790 & -3673669570 & -9270547450 & -12163627347 & -8332648800 & -2428876217 & 0 & 60439480 & 3015890 & 15504 & -31008 & -6031780 & -120878960 & 0 & 4857752434 & 16665297600 \\
14 & -11619550320 & -86026550610 & -276775252450 & -495205128390 & -520780842570 & -307670159230 & -81976263336 & 0 & 2428876217 & 182157600 & 2568819 & -5137638 & -364315200 & -4857752434 & 0 & 163952526672 \\
15 & -181746027600 & -1587521062470 & -6209911975410 & -14096056649296 & -20076314098740 & -18017053206978 & -9545387953870 & -2386587310945 & 0 & 81976263336 & 8332648800 & 213810173 & -429868426 & -16665297600 & -163952526672 & 0 \\
\hline
\end{tabular}
}
\caption{The coefficients $b_{i,j}$ in \eqref{eq:5d_1inst} for $0\le i,j \le 15$.}
\end{table}

Finally, we compare our result of the normalized elliptic genus $\widetilde{\mb{E}}_1(v,q)$ with the generalized Schur index calculated in \cite{Deb:2025ypl}. To do so, we extract the function $L_G(v,q)$ as defined in \cite[eq (7.29)]{DelZotto:2016pvm} from $\widetilde{\mb{E}}_1(v,q)$. We have for $G=\Ehalf$,
\be\ba
    L_G(v,q)&=(1+190 v^2+15504 v^4+749360 v^6+24732110 v^8+605537790 v^{10}+11619550320 v^{12}+\cdots)\\
    &+q(191+33650 v^2+2568819 v^4+117126070 v^6+3673669570 v^8+86026550610 v^{10}+\cdots)\\
    &+q^2(18526+3015890 v^2+215308893 v^4+9270547450 v^6+276775252450 v^8+6209911975410 v^{10}+\cdots)\\
    &+q^3(1214577+182157600 v^2+12163627347 v^4+495205128390 v^6+14096056649296 v^8+\cdots)+\mathcal{O}(q^4)\,.
\ea\ee
In the unflavored limit $v=q^{1/4}$, $L_G(q^{1/4},q)$ recovers the generalized Schur index and the level $-5$ character for the $\Ehalf$ VOA as we have presented in \eqref{eq:LG}.

\end{document}